\def\papertitle{Sample Rate Independent Recurrent Neural Networks for Audio Effects Processing}
\def\paperauthorA{Alistair Carson}
\def\paperauthorB{Alec Wright}
\def\paperauthorC{Jatin Chowdhury}
\def\paperauthorD{Vesa Välimäki}
\def\paperauthorE{Stefan Bilbao}

\documentclass[twoside,a4paper]{article}
\usepackage{etoolbox}
\usepackage{siunitx}
\usepackage{dafx_24}
\usepackage[hyphens]{url}

\usepackage{amsmath,amssymb,amsfonts,amsthm}
\usepackage{euscript}
\usepackage[T1]{fontenc}
\usepackage[utf8]{inputenc}
\usepackage{ifpdf}
\usepackage[english]{babel}
\usepackage{caption}
\usepackage{subfigure} 
\usepackage{color}

\usepackage{xcolor}
\def\SB[#1]{\textcolor{blue}{#1}}
\def\Vesa[#1]{\textcolor{green}{#1}}
\def\AC[#1]{\textcolor{orange}{#1}}
\def\Jatin[#1]{\textcolor{cyan}{#1}}
\input glyphtounicode
\ninept

\newcounter{numauth}
\newcounter{listcnt}
\newcommand\authcnt[1]{\ifdefined#1 \stepcounter{numauth} \fi}

\newcommand\addauth[1]{
\ifdefined#1 
\stepcounter{listcnt}
\ifnum \value{listcnt}<\value{numauth}
\appto\authorslist{, #1}
\else
\appto\authorslist{~and~#1}
\fi
\fi}
\authcnt{\paperauthorB}
\authcnt{\paperauthorC}
\authcnt{\paperauthorD}
\authcnt{\paperauthorE}
\authcnt{\paperauthorF}
\authcnt{\paperauthorG}
\authcnt{\paperauthorH}
\authcnt{\paperauthorI}
\authcnt{\paperauthorJ}
\def\authorslist{\paperauthorA}
\addauth{\paperauthorB}
\addauth{\paperauthorC}
\addauth{\paperauthorD}
\addauth{\paperauthorE}
\addauth{\paperauthorF}
\addauth{\paperauthorG}
\addauth{\paperauthorH}
\addauth{\paperauthorI}
\addauth{\paperauthorJ}

\usepackage{times}

\newif\ifpdf
\ifx\pdfoutput\relax
\else
   \ifcase\pdfoutput
      \pdffalse
   \else
      \pdftrue
\fi

\ifpdf 
  \usepackage[pdftex,
    pdftitle={\papertitle},
    pdfauthor={\authorslist},
    pdfsubject={Proceedings of the 27th International Conference on Digital Audio Effects (DAFx24)},
    colorlinks=false, 
    bookmarksnumbered, 
    pdfstartview=XYZ 
  ]{hyperref}
  \usepackage[pdftex]{graphicx}
\else 
  \usepackage[dvips]{epsfig,graphicx}
  \usepackage[dvips,
    pdftitle={\papertitle},
    pdfauthor={\authorslist},
    pdfsubject={Proceedings of the 27th International Conference on Digital Audio Effects (DAFx24)},
    colorlinks=false, 
    bookmarksnumbered, 
    pdfstartview=XYZ 
  ]{hyperref}
\fi
\usepackage[hypcap=true]{caption}
\title{\papertitle}

\affiliation{
\paperauthorA $^{1}$\sthanks{A. Carson is funded by the Scottish Graduate School of Arts and Humanities (SGSAH).}\hspace{-2pt},
\paperauthorB $^{1}$,
\paperauthorC $^{2}$,
\paperauthorD $^{3}$,
\paperauthorE $^{1}$
}
{
$^1$\href{https://www.acoustics.ed.ac.uk}{Acoustics and Audio Group}, University of Edinburgh, Edinburgh, United Kingdom\\
$^2$ \href{https://chowdsp.com/}{Chowdhury DSP}, Sammamish, WA, USA\\
$^3${\href{https://www.aalto.fi/en/aalto-acoustics-lab}{Acoustics Lab}, Department of Information and Communications Engineering, Aalto University, Espoo, Finland \\
{\tt \href{mailto:alistair.carson@ed.ac.uk}{alistair.carson@ed.ac.uk}
}
}
}

\begin{document}
\ifpdf 
  \DeclareGraphicsExtensions{.png,.jpg,.pdf}
\else  
  \DeclareGraphicsExtensions{.eps}
\fi


\maketitle

\begin{abstract}
In recent years, machine learning approaches to modelling guitar amplifiers and effects pedals have been widely investigated and have become standard practice in some consumer products. In particular, recurrent neural networks (RNNs) are a popular choice for modelling non-linear devices such as vacuum tube amplifiers and distortion circuitry. One limitation of such models is that they are trained on audio at a specific sample rate and therefore give unreliable results when operating at another rate. Here, we investigate several methods of modifying RNN structures to make them approximately sample rate independent, with a focus on oversampling. In the case of integer oversampling, we demonstrate that a previously proposed delay-based approach provides high fidelity sample rate conversion whilst additionally reducing aliasing. For non-integer sample rate adjustment, we propose two novel methods and show that one of these, based on cubic Lagrange interpolation of a delay-line, provides a significant improvement over existing methods. To our knowledge, this work provides the first in-depth study into this problem. 
\end{abstract}

\section{Introduction}

Virtual Analog (VA) modelling \cite{DAFXVAchapter} refers to the digital emulation of analog systems, such as guitar amplifiers and distortion effects. The aim is to replace bulky and expensive hardware with software, typically implemented as an audio plugin, to be used in a digital audio workstation.

Approaches to VA modelling are generally divided between {\it white-box} methods \cite{Karjalainen06, Paiva12}, which derive equations based on circuit modelling, and {\it black-box} methods, which define a general model structure and use data to optimise the model parameters \cite{novak2010chebyshev, orcioni2018identification}. Black-box models based on neural networks have become a popular approach to creating digital models of guitar amplifiers and distortion effects in recent years, including models based on Convolutional Neural Networks \cite{Damskaag2018}, Recurrent Neural Networks (RNNs) \cite{Parker2019, Wright2019RNN} and Differentiable Digital Signal Processing \cite{Nercessian:ICASSP2021}. 

Audio plugins are frequently implemented at multiple sample rates (SR), allowing the user to choose. For white-box methods, adjusting the SR is straightforward, as it is generally explicitly included as a parameter in the model. For neural network methods, the SR of the model is fixed to that of the training data and is generally not adjustable after training. The exception to this is the State-Trajectory Network (STN) \cite{Parker2019}, which has an adjustable SR but also requires additional state information to be recorded from the target device.

In this paper, we focus on a particular class of RNN model that is commonly applied to guitar amplifier modelling \cite{Wright2019RNN}. This model can achieve excellent perceptual quality at a relatively low computational cost \cite{Wright2020, ndsp2023, cassidy2023perceptual}. Specifically, we investigate some previously proposed methods and two novel extensions for oversampling the models at inference-time. We show that with the correct choice of method, the SR of a pre-trained RNN model can be adjusted with very little impact on the emulation quality, whilst additionally reducing aliasing. For guitar amplifier plugins based on RNNs, this allows for oversampling without requiring multiple models to be trained at different SRs. Furthermore, it allows the processing of signals with arbitrary SRs without the need for resampling at run-time.

This paper is outlined as follows: Sec.~\ref{sec:problem} formally outlines the problem statement with Sec. \ref{sec:methods} presenting the possible methods of modifying RNNs to be approximately SR independent. In Sec. \ref{sec:linear}, the methods are evaluated for a simplified test problem of a linear filter, and in Sec. \ref{sec:neural_nets} evaluation is carried out on several neural network models of guitar amplifiers and effects pedals. Sec. \ref{sec:conc} provides concluding remarks and areas of further work.

\section{Problem statement} \label{sec:problem}
Consider a continuous-time input audio signal $x(t)$ that has been sampled at a rate of $F_s=1/T$ to give $x^n \equiv x[n] \approx x(nT)$, where $n$ is an integer sample index. In this work we consider recurrent neural networks (RNNs) of the form:
\begin{subequations}\label{eq:rnn_and_fc}
\begin{align}
    \mathbf{h}^{n} &= f\left(\mathbf{h}^{n-1}, x^n \right) \label{eq:rnn} \\ 
    y^n   &= g \left( \mathbf{h}^n, x^n \right),\label{eq:fc}
\end{align}
\end{subequations}
where $\mathbf{h}^n \in \mathbb{R}^{H \times 1}$ is the hidden state of length $H$ and $y^n$ is the output signal. This class of model has been extensively used in recent years for modelling guitar amplifiers and effects pedals \cite{Wright2020}. The recurrent layer, Eq.~\eqref{eq:rnn}, is typically a Gated Recurrent Unit (GRU) or a Long Short-Term Memory Network (LSTM), but this work is not tied to any specific recurrent unit architecture. The function $f(\cdot)$ is in general non-linear and typically a combination of hyperbolic tangent and logistic activation functions. Alternatively, Eq.~\eqref{eq:rnn} can be interpreted more generally as a non-linear auto-regressive model (NARMAX) \cite{BillingsNARX}. Here we take $g(\cdot)$ to be a fully-connected memoryless affine layer, which is by definition SR independent, therefore do not consider further its influence when operating at different SRs.

In this work we consider the case where the RNN has been trained on audio signals at a certain SR, $F_s$, but at inference we wish to operate at an oversampled rate of $F_s' = 1/T' = M F_s$, where $\{M \in \mathbb{R} \,| \,M > 1\}$ is the integer or non-integer oversampling factor. Assuming the input signal resampled at the new rate, so that $\tilde{x}^n \approx x(n T')$, we seek a model which correctly predicts the oversampled hidden state time-series, $\tilde{\bf h}^n$,
such that the perceptual quality of the output signal $\tilde{y}^n$ is not negatively impacted by the change of SR. We would expect that oversampling reduces aliasing artifacts caused by the non-linear activations -- which could in fact improve the perceived quality -- therefore we seek an oversampling method that reduces aliasing without otherwise affecting the original harmonic spectral content.

In this work we focus on a base SR of $F_s=$ \SI{44.1}{\kHz} and consider oversampling relative to this. SR conversion from a higher to a lower rate is left to further work. We evaluate the quality of models based on their accuracy within the original Nyquist limit, i.e. we only consider frequency content below \SI{22.05}{\kHz} and consider potential artifacts beyond this negligible due to inaudibility. 

\begin{figure}[t]
\centerline{\includegraphics[scale=0.6]{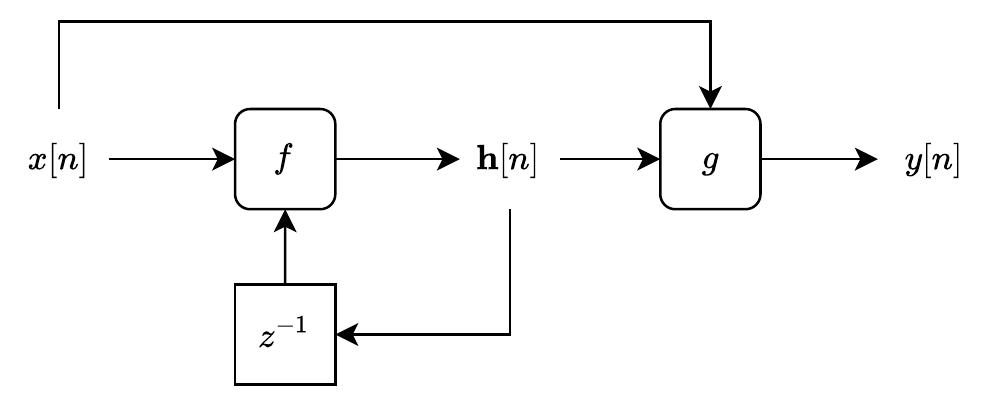}}
\caption{\label{fig:lidl}{\it Baseline recurrent neural network architecture studied in this work, where $f$ is a non-linear recurrent cell and $g$ is a fully connected affine layer.}} 
\end{figure}

\section{Methods}\label{sec:methods}
In this section we outline the existing approaches to SR independent RNNs, as well as a proposed extension to these. For comparison, for a given $f$ and $g$, we additionally consider computing the same recursion, Eq.~\eqref{eq:rnn}, at a different SR, and refer to this as the ``naive'' method.

\subsection{State-trajectory network method (STN)}\label{subsec:stn}
\begin{figure}[t!]
\centerline{\includegraphics[scale=0.7]{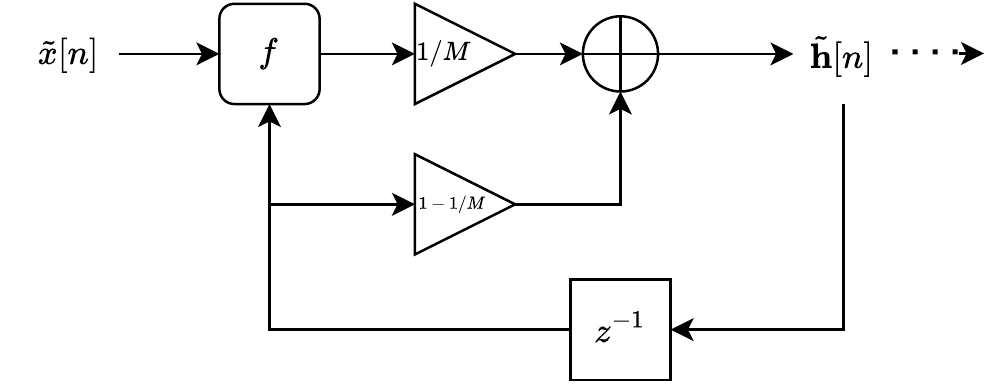}}
\caption{\label{fig:stn}{\it STN method: modified RNN architecture for changing the SR by factor of $M$ by scaling the state residual \cite{Parker2019}.}}
\end{figure}

Recent work has explored the connection between RNNs and numerical solvers for state-space systems \cite{Parker2019, Peussa2021, Wilczek2022}. In this section we show that interpreting an RNN in this way provides one possible method of SR modification, and will refer to it as the State-trajectory network (STN) method after the work in which it was originally proposed \cite{Parker2019}. The general RNN functional form Eq.~\eqref{eq:rnn} can be rewritten in terms of the state residual, $r({\bf h}^{n-1}, x^{n})$:
\begin{align}
    \mathbf{h}^{n} &= {\bf h}^{n-1} + \underbrace{\left(-{\bf h}^{n-1}+f\left(\mathbf{h}^{n-1}, x^n \right)\right)}_{r({\bf h}^{n-1}, x^{n})}\label{eq:residual} 
\end{align}
This can now be viewed as the Forward Euler (FE) numerical solution to the ordinary differential equation (ODE)
\begin{equation}
    \dot{{\bf h}} = F_{s} r \left(\mathbf{h}, x \right)
\end{equation}
where ${\bf h}^{n}$ is assumed drawn from an underlying continuous state variable ${\bf h}(t)$ at $t=nT$, and $x^{n}$ is advanced by one time step. It follows that simulating this ODE at a new SR $F_{s}'$ leads to
\begin{equation}\label{eq:stn}
\begin{aligned}
        \tilde{\mathbf{h}}^{n} &= \tilde{{\bf h}}^{n-1}+\frac{1}{M} r \left(\tilde{\mathbf{h}}^{n-1}, \tilde{x}^n \right)  \\
        &= \left( 1 - \frac{1}{M} \right)  \tilde{\mathbf{h}}^{n-1} + \frac{1}{M}{f}\left(\tilde{\mathbf{h}}^{n-1}, \tilde{x}^n \right) 
\end{aligned}
\end{equation}
where $M = F'_s / F_s$ is the oversampling factor. Parker et al. \cite{Parker2019} investigated modelling non-linear state-space systems with a model of this form, where they trained a multi-layer perceptron in place of the function $r$. In their work they suggest that the factor $M$ could be used to alter the time-scale of the recursion, with some caveats: for $M > 1$, the system may lack high frequency behaviour and for $M < 1$ that aliasing might occur \cite{Parker2019}. In this work we investigate the former and show that SR conversion in this manner does indeed lead to a low-passing effect within the audible spectrum. We refer to predicting the hidden states via Eq.~\eqref{eq:stn} as the STN method of SR modification. The corresponding block diagram is shown in Figure \ref{fig:stn}.
\subsection{Linearly interpolated delay-line (LIDL)}
An alternative method is to modify the delay length (in samples) in the original RNN architecture, as originally proposed in \cite{Chowdhury2022}. Consider the RNN in Eq. \eqref{eq:rnn} as approximating a continuous time system, which implements a fixed delay of $T = 1/F_s$ s:
\begin{equation}
    \mathbf{h}(t) = f \left( \mathbf{h}(t - T), x(t)\right),
\end{equation}
This system can be sampled with another time-step, $T'$ by setting $t = n T'$ to obtain:
\begin{align}
    \mathbf{h}(nT') = f \left( \mathbf{h}((n - M)T'), x(nT')\right)  
\end{align}
where $M$ is as defined in the previous section. We can then approximate this system in discrete time as:
\begin{equation}\label{dline}
    \tilde{\mathbf{h}}^{n} = f\left(\tilde{\mathbf{h}}^{n-M}, \tilde{x}^{n} \right)
\end{equation}
where $\tilde{\mathbf{h}}^n \equiv \tilde{\mathbf{h}}[n] \approx \mathbf{h}(nT')$. In this method, the hidden state of the RNN is delayed by the same duration (in s) as when operating at the original SR. For integer oversampling, $M \in \mathbb{Z}^{+}$, and the delay of $M$ samples can be easily implemented with a delay line.

However, in the case of non-integer resampling (e.g.~from \SI{44.1}{\kHz} to \SI{48}{\kHz}) the state at time step $n - M$ is not defined. In this case, Chowdhury \cite{Chowdhury2022} proposed using linear interpolation between the two nearest states to approximate the unknown state:
\begin{eqnarray}\label{linterp}
    \tilde{{\bf h}}^{n - M} \approx (1 - \Delta) \tilde{\mathbf{h}}^{n- \lfloor M \rfloor} + \Delta \tilde{\mathbf{h}}^{n- \lfloor M \rfloor - 1}
\end{eqnarray}
where $\Delta = M - \lfloor M \rfloor$. Linear interpolation is known to be a good approximation for very small fractional delays due to its linear phase response \cite{Valimaki2000}, but shows increasing errors in magnitude at higher frequencies. It is worth noting that for $M<1$, Eq. \eqref{linterp} becomes implicit, therefore this method is not suitable for lowering the SR. In theory, one-sided extrapolation could be used here instead, but this is left to further work. 

\begin{figure}[t!]
\centerline{\includegraphics[scale=0.7]{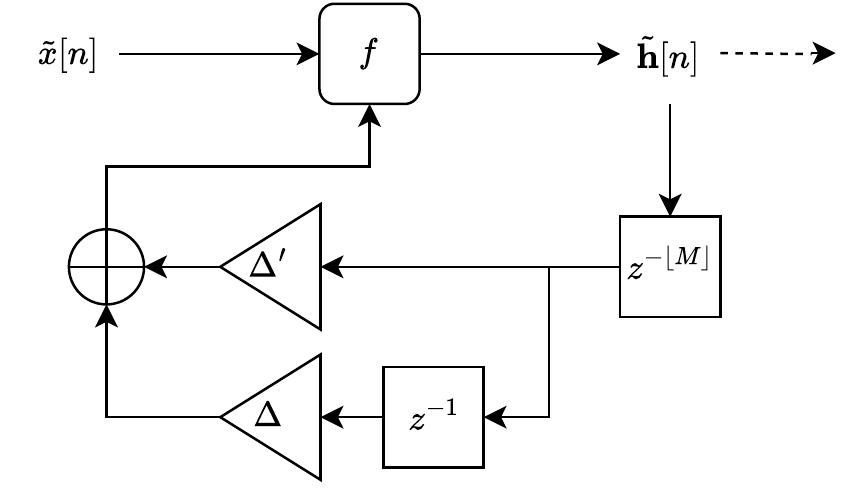}}
\caption{\label{fig:lidl}{\it LIDL method: modified RNN architecture for oversampling by a factor of $M$ by linearly interpolating between successive states. $\Delta = M - \lfloor M \rfloor$ and $\Delta' = 1 - \Delta$. Adapted from \cite{Chowdhury2022}.}}
\end{figure}

\subsection{All-pass delay line method (APDL)}
\begin{figure}[ht]
\centerline{\includegraphics[scale=0.7]{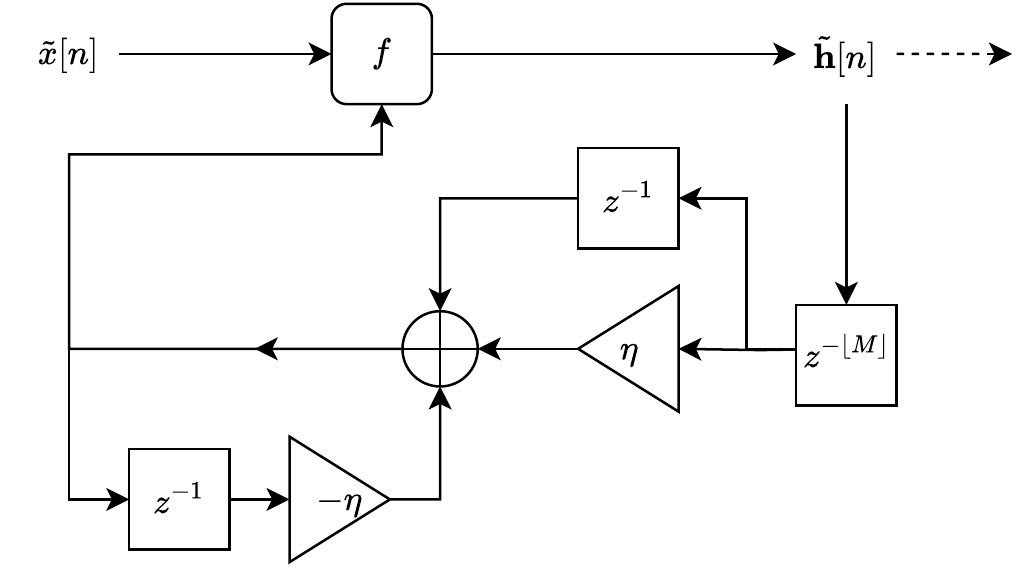}}
\caption{\label{fig:apdl}{\it Proposed APDL method: modified RNN architecture for oversampling by a non-integer factor of $M$ with an all-pass filter that implements a fractional delay. $\eta$ is given by Eq.~\eqref{ap_param}.}}
\end{figure}

The LIDL method can be interpreted as using a first-order FIR filter to approximate the fractional delay when oversampling by a non-integer factor. In this work we also consider an IIR filter approach to this by embedding an all-pass filter in the state feedback loop, as shown in Figure \ref{fig:apdl}. The first-order all-pass filter has the transfer function:
\begin{equation}\label{apf}
    H_{ \eta }(z) = \frac{\eta + z^{-1}}{1 + \eta z^{-1}},
\end{equation}
and is known to approximate a delay of $\Delta$ samples when \cite{Laakso1996}:
\begin{equation}\label{ap_param}
    \eta = \frac{1 - \Delta }{1 + \Delta }.
\end{equation}
This approximation is only valid in the low-frequency limit, however. Unlike linear interpolation, all-pass filter interpolation gives a unity magnitude response but shows increasing errors in phase at high frequencies \cite{Laakso1996, Valimaki2000}. 
For an oversampling ratio of $M$, a fractional delay of $\Delta = M - \lfloor M \rfloor $ samples is required, so from Eq.~\eqref{ap_param}, the following recursion results:
\begin{equation}
    \mathbf{a}^n = \eta \cdot (\tilde{\mathbf{h}}^{n-\lfloor M \rfloor} - \mathbf{a}^{n-1}) + \tilde{\mathbf{h}}^{n-\lfloor M \rfloor -1}, 
\end{equation}
where $\mathbf{a} \in \mathbb{R}^{N \times 1}$ is the all-pass filter state. The current all-pass filter state can then be substituted into Eq. \eqref{dline} as an approximation to $\tilde{\mathbf{h}}^{n - M}$. 

Note that for integer $M$, the APDL method and the LIDL method are identical. Likewise, the APDL method is not suitable for lowering the SR ($M < 1$). Furthermore, as $\Delta \to 0$, the filter relies on pole-zero cancellation to give a flat frequency response, which can introduce numerical errors when operating with finite-precision arithmetic. As such, it has been previously recommended to design APF interpolators such that, where possible, $\Delta \in [0.1, 1.1]$ \cite{PASPWEB2010}. In Section \ref{sec:neural_nets} we show that this does indeed cause problems for SR conversion from \SI{44.1}{\kHz} to \SI{48}{\kHz}.

\subsection{Cubic Lagrange interpolated delay-line (CIDL)}
The final method we consider involves using a higher order FIR filter to approximate the fractional delay. In this work we propose using a cubic Lagrange interpolation \cite{Laakso1996}, where the filter order is $K=3$. Higher order filters were not investigated but are left to further work. In general, a fractional delay of $\Delta$ samples can be achieved by convolution with the impulse response \cite{Laakso1996}:
\begin{eqnarray}\label{eq:lagrange_coeffs}
    l_{\Delta}[n] = \prod_{k=0, k \neq n}^{K} \frac{\Delta - k}{n - k}, \quad n = 0, \dots, K.
\end{eqnarray}
For an oversampling factor of $M$, we can convolve this kernel with $K+1$ previous states to approximate the state at time-step $n-M$:
\begin{eqnarray}
    \tilde{\bf h}^{n - M} \approx \sum_{k=0}^{K} l_{\Delta}[k] \cdot \tilde{\bf h}^{n - k - \gamma}.
\end{eqnarray}
where for $M>1$:
\begin{align}\label{eq:gamma_delta}
    \gamma = 1+\lfloor |M-2|\rfloor, \quad \Delta = M-\lfloor|M-2|\rfloor-1.
\end{align}
Note that for $M < 2$ this results in an asymmetric selection of nearest neighbouring samples over which to interpolate. In general this will generate more error than a centered design but is unavoidable due to the state at time-step $n$ being unknown. For $M>2$, a centered design is always realisable and therefore recommended. 
\begin{figure}[ht]
\centerline{\includegraphics[scale=0.7]{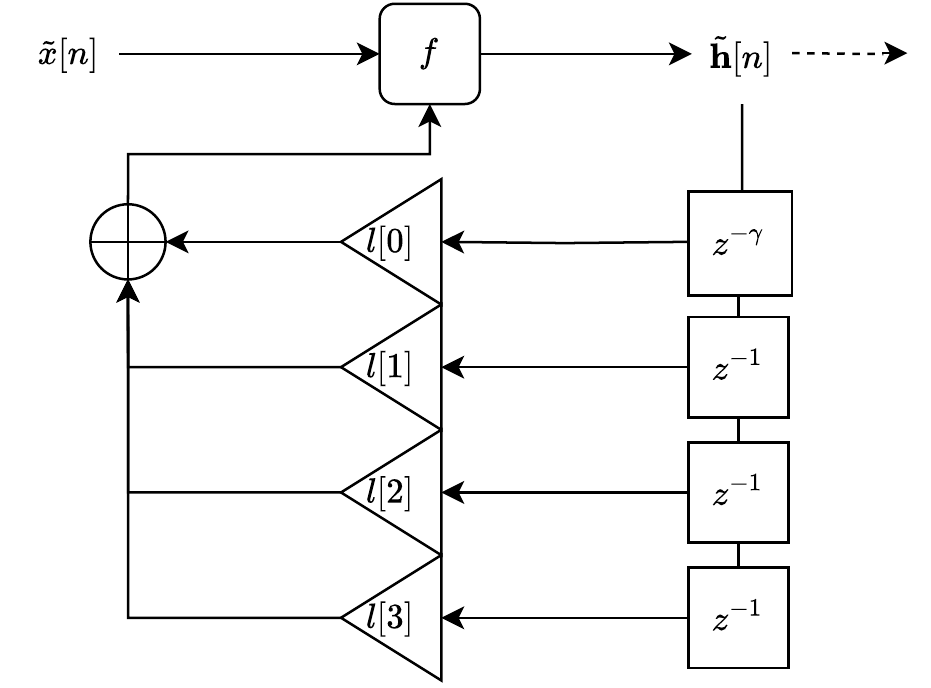}}
\caption{\label{fig:cidl}{\it Proposed CIDL method: modified RNN architecture for oversampling by a factor of $M$ with third-order Lagrange delay-line interpolation. The filter coefficients, $l$, and delay $\gamma$  are given by Eqs.~\eqref{eq:lagrange_coeffs} and \eqref{eq:gamma_delta} respectively.}}
\end{figure}

\subsection{Computational cost}
A software benchmark was constructed, with a series
of GRU networks with hidden sizes ranging from 24 to
72, trained at \SI{44.1}{\kHz} SR. The networks were
implemented in \texttt{C++}, using the
\texttt{RTNeural} library \cite{chowdhury2021rtneural},
and used to process $T_{{\rm audio}}=100$ s of audio at
\SI{96}{\kHz} SR, using each of the SR adaptation methods
described above. The processing time $T_{{\rm proc}}$ for
each network and adaptation method was measured, and
used to compute the ``real-time factor''
$F_{{\rm RT}} = T_{{\rm audio}} / T_{{\rm proc}}$. The performance
benchmark was run on a 2021 Apple M1 MacBook Pro.

Table \ref{tab:bench_results} shows the average
percent change in real-time factor across the
different network sizes, relative to the naive 
implementation. The delay and STN methods tended
to perform the best, having almost no impact on
the measured real-time factor, while the
interpolation-based methods were slightly slower.
These results are expected, as the
interpolation-based methods (particularly CIDL
and APDL) require more computational work than
the other methods. However, the additional cost of 
performing SR adaptation with any of the proposed
methods is small relative to the computational cost
of the GRU network overall.

\begin{table}[t!]
    \centering
    \begin{tabular}{|c|c|c|c|c|}
    \hline
    Delay & STN & LIDL & APDL & CIDL  \\
    \hline
    -0.56\% & -0.57\% & -2.13\% & -3.24\% & -3.17\%  \\
    \hline
    \end{tabular}
    \caption{\label{tab:bench_results} \itshape The
    average change to the real-time factor measured
    in the performance benchmarks across networks
    of varying size, relative to the naive
    implementation.}
\end{table}

\begin{figure*}[h!]
\centerline{\includegraphics[width=\textwidth, clip, trim=0mm 0mm 0mm 0mm]{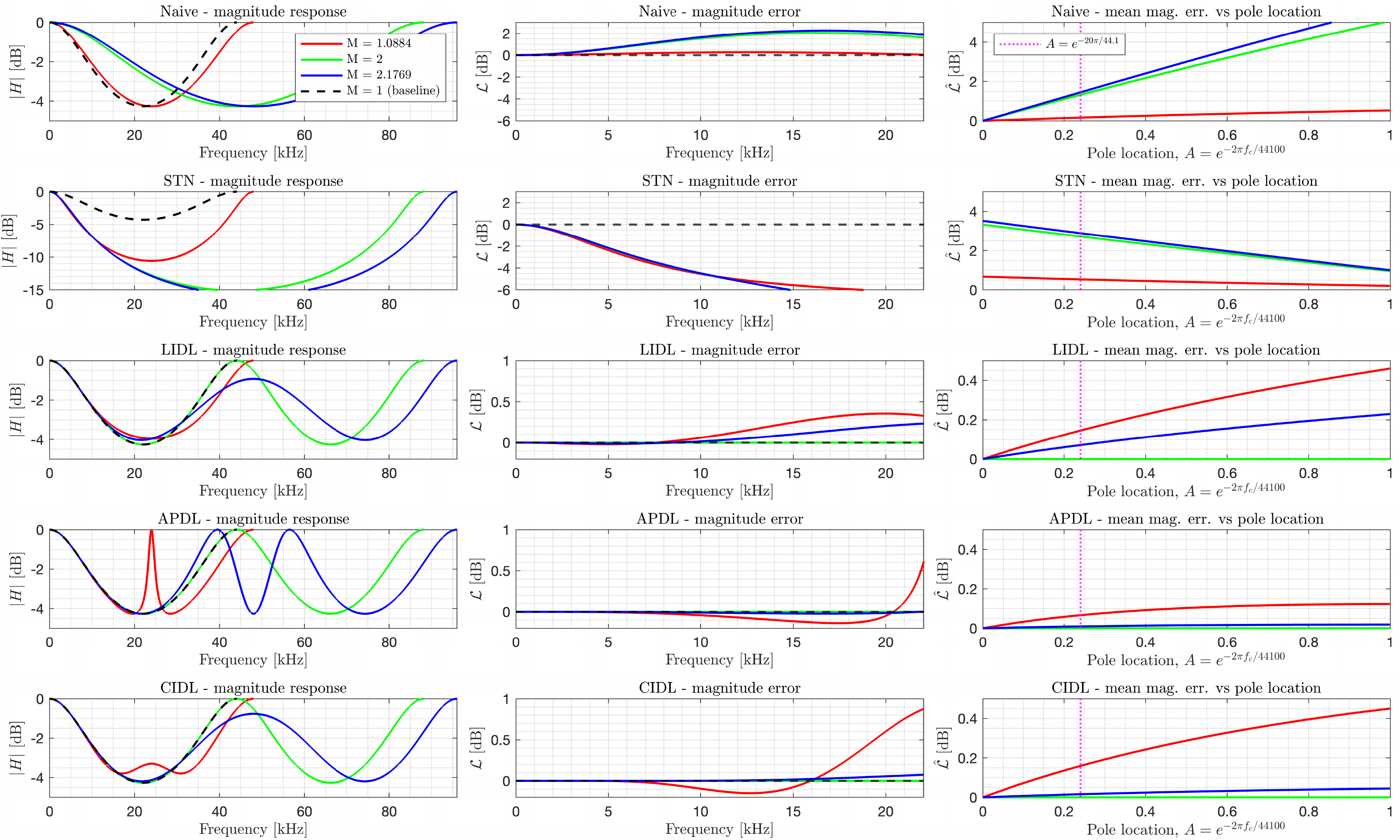}}
\caption{\label{fig:linear_subplots}{\it Left: magnitude responses of the modified and unmodified one-pole filter \eqref{eq:one_pole_naive} - \eqref{eq:one_pole_cidl} for different oversampling factors $M$. Note that a single cycle of the full-band DFT is shown for all SRs. Centre: the corresponding spectral error \eqref{eq:spec_error} when compared to the unmodified system at the base rate, $F_s= $ \SI{44.1}{\kHz}. The frequency axis is cropped to the original Nyquist range. Right: mean spectral error (over original Nyquist range) against target system pole location, $A = e^{-2 \pi f_c / F_s}$. In LH and centre columns, the pole location is fixed at $A = e^{-20 \pi / 44.1}$; in RH column this point is indicated by the magenta dotted line. NB y-axis limits differ across subplots.}}
\end{figure*}
\section{Linear analysis} \label{sec:linear}
In this section, we consider a simplified problem of replacing system Eq.~\eqref{eq:rnn} with a one-pole linear filter of the form:
\begin{equation}\label{eq:one_pole}
    h^{n} = Ah^{n-1} + x^{n}, \quad y^n = h^n,
\end{equation}
where $A = e^{-2 \pi f_c/F_s}$. This can be interpreted as a discretization (which is exact under zero-input conditions) of a one-pole RC filter with cutoff frequency, $f_c$ and sampled at $F_s$, driven by some discrete-time signal $x^n$. In neural network terminology, this could be interpreted as single-weight RNN with linear activation. The task of operating the recursion of Eq.~\eqref{eq:one_pole} at arbitrary SRs is trivial: the parameter $A$ can be computed for any $F_s$. However, we consider the case where $A$ is \textit{fixed} and apply the proposed methods to operate at different SRs. This is a realistic introductory problem to SR independent RNNs where the model weights are not parameterized in terms of SR, therefore we are forced to look at algorithmic rather than parametric changes to the model.

In the following analysis, we apply the methods outlined in Sec.~\ref{sec:methods} to the one-pole IIR filter and for different oversampling ratios $M$ compare the frequency response to the original, unmodified system at the base SR. The linearity of both the system and methods allow the analysis to be carried out in the frequency domain. The baseline frequency response, corresponding to the system operating at the base SR, $F_s = 1/ T$ can be derived analytically:
\begin{equation}\label{eq:one_pole_tf}
    H_{{\rm BASE}}(\omega) = \frac{1}{1 - Ae^{-j \omega T}},
\end{equation}
where $\omega$ is angular frequency in rad / s. The respective frequency responses of the system modified with the studied methods can then be derived:
\begin{subequations}
\begin{gather}
        H_{\rm NAIVE}(\omega) = \frac{1}{1 - Ae^{-j \omega T'}} \label{eq:one_pole_naive} \\
        H_{{\rm STN}}(\omega) = \frac{1}{M - \left(M + A - 1\right)e^{-j \omega T'}} \label{eq:one_pole_tf_stn} \\
        H_{\rm LIDL}(\omega) = \frac{1}{1 - A \left( 1 - \alpha + \alpha e^{-j \omega T'} \right) e^{-j \omega T' \lfloor M \rfloor}}  \label{eq:one_pole_tf_lidl} \\
        H_{\rm APDL}(\omega) = \frac{1 + \eta e^{-j \omega T'}}{1 + \eta e^{-j \omega T'} - A(\eta + e^{-j \omega T'})e^{-j \omega T' \lfloor M \rfloor} } \label{eq:one_pole_apdl} \\
        H_{\rm CIDL}(\omega) = \frac{1}{1 - A \sum_{k=0}^{k=3}l_{\Delta}[k]e^{-j \omega T'(k + \gamma)}} \label{eq:one_pole_cidl}
\end{gather}
\end{subequations}
where $M$ is the oversampling factor, $T' = T / M$ is the operating sample period and $\Delta$, $\gamma$ are defined in Eq.~\eqref{eq:gamma_delta}. 

Figure \ref{fig:linear_subplots} (left) shows the frequency responses \eqref{eq:one_pole_naive} -- \eqref{eq:one_pole_cidl} for different oversampling factors. The baseline frequency response, $H_{\rm BASE}$ is overlaid on each subplot as a black dashed line. Figure \ref{fig:linear_subplots} (center) shows spectral error magnitude as a function of frequency, in decibels, defined as:
\begin{equation}\label{eq:spec_error}
    \mathcal{L}(\omega) = 20 \log_{10} \left| \frac{H_{\rm <method>}(\omega)}{H_{\rm BASE}(\omega)} \right|.
\end{equation}
In Figure \ref{fig:linear_subplots} (right) the y-axis shows the mean spectral error, $\hat{\mathcal{L}}$, determined by averaging Eq.~\eqref{eq:spec_error} over the frequency in the range $\omega \in (0, F_s/2)$ where $F_s/2$ is the Nyquist limit at the base SR. The x-axis is the pole-location $A$, in the (unmodified) IIR filter \eqref{eq:one_pole}.

\subsection{Integer oversampling}
For integer oversampling, the delay-based methods (LIDL, APDL, and CIDL) all reduce to the same frequency response which is identical to the original frequency response. This comes at the cost of frequency aliasing above the original SR. This can be seen in Figure \ref{fig:linear_subplots} (left): when operating at an oversampled rate of $M=2$, the original spectrum is mirrored about the original SR. In the context of a low-pass filter, this may seem like a major problem due to the creation of fictitious high-frequency behaviour. However, assuming the input signal, $x$, is bandlimited to the original Nyquist limit then this high frequency behaviour will not be present in the output signal. This assumption is reasonable in the context of oversampling for aliasing reduction, given an appropriate choice of up-sampling filter with sufficient stop-band attenuation \cite{Kahles2019}.

In contrast to the delay-line methods, the STN method does not exhibit any frequency-mirroring, but has a low-passing effect on the frequency response. Although not pictured in the plot, as $F'_s \to \infty$, the frequency response converges, which is intuitive given its resemblance to an explicit numerical solver. However it converges to the incorrect response therefore represents a different system to the original. The naive method simply stretches the function along the frequency axis, causing significant frequency warping below the original Nyquist limit.

\subsection{Non-integer oversampling}
Non-integer oversampling exposes differences between the LIDL, APDL and CIDL methods. In Figure \ref{fig:linear_subplots} (center), the LIDL method acts as a high-shelving filter within the original Nyquist range. At \SI{48}{\kHz} ($M=1.0884$), the APDL provides a flatter response except at very high frequencies, where the pole at the new Nyquist frequency causes a resonance. The CIDL method gives a different error profile to the LIDL method but the mean spectral error is approximately the same. At \SI{96}{\kHz}, both APDL and CIDL provide a significantly flatter response than LIDL. Note that Figure \ref{fig:linear_subplots} (center) shows the magnitude error for a RC filter at one particular cut-off frequency only ($f_c$ = \SI{10}{\kHz}).

To get a broader view, we can compare the methods across the full range of possible pole locations. Figure \ref{fig:linear_subplots} (right) shows these results. In this case, the absolute spectral error has been averaged over the Nyquist range of the base SR. It can be seen that, at \SI{48}{\kHz}, the LIDL and CIDL methods give similar results, but at \SI{96}{\kHz} the higher-order interpolation (CIDL) gives a notable improvement particularly as the system pole approaches the unit circle pole, i.e. as $f_c \to 0$. In the other extreme, as $f_c \to \infty$, the error of all three delay-based methods approaches zero, which is reassuring as in this limit the RC filter should pass all frequencies with no attenuation. In contrast, the STN method gets worse in this limit with an error of up to 3\,dB. At both non-integer oversampling factors the APDL gives the lowest spectral error across the full range of pole locations. This suggests that APDL is the best method for non-integer SR conversion -- a result which is perhaps not surprising given that the APF has, by definition, a flat magnitude response. However, we will see in Sec.~\ref{sec:neural_nets} that this result does not always translate to good behaviour for more complex, nonlinear systems, and that the FIR-based methods may indeed be preferable.

\section{Evaluation on Neural Networks}\label{sec:neural_nets}
The different methods were evaluated on 18 pre-trained LSTM networks found in the GuitarML Tone Library \footnote{\href{https://guitarml.com/tonelibrary/tonelib-pro.html}{https://guitarml.com/tonelibrary/tonelib-pro.html}}. These are models of various effects including low- and high-gain amps, fuzz pedals,  acoustic simulation and one compressor pedal. The models are designed for a plug-in operating at \SI{44.1}{\kHz} -- so we assume the models were correctly trained on audio at this SR. We make our testing code publicly available and provide audio examples \footnote{\href{https://a-carson.github.io/dafx24_sr_indie_rnn/}{https://a-carson.github.io/dafx24\_sr\_indie\_rnn/}}.
\begin{figure*}[h!]
    \centerline{\includegraphics[width=.99\textwidth]{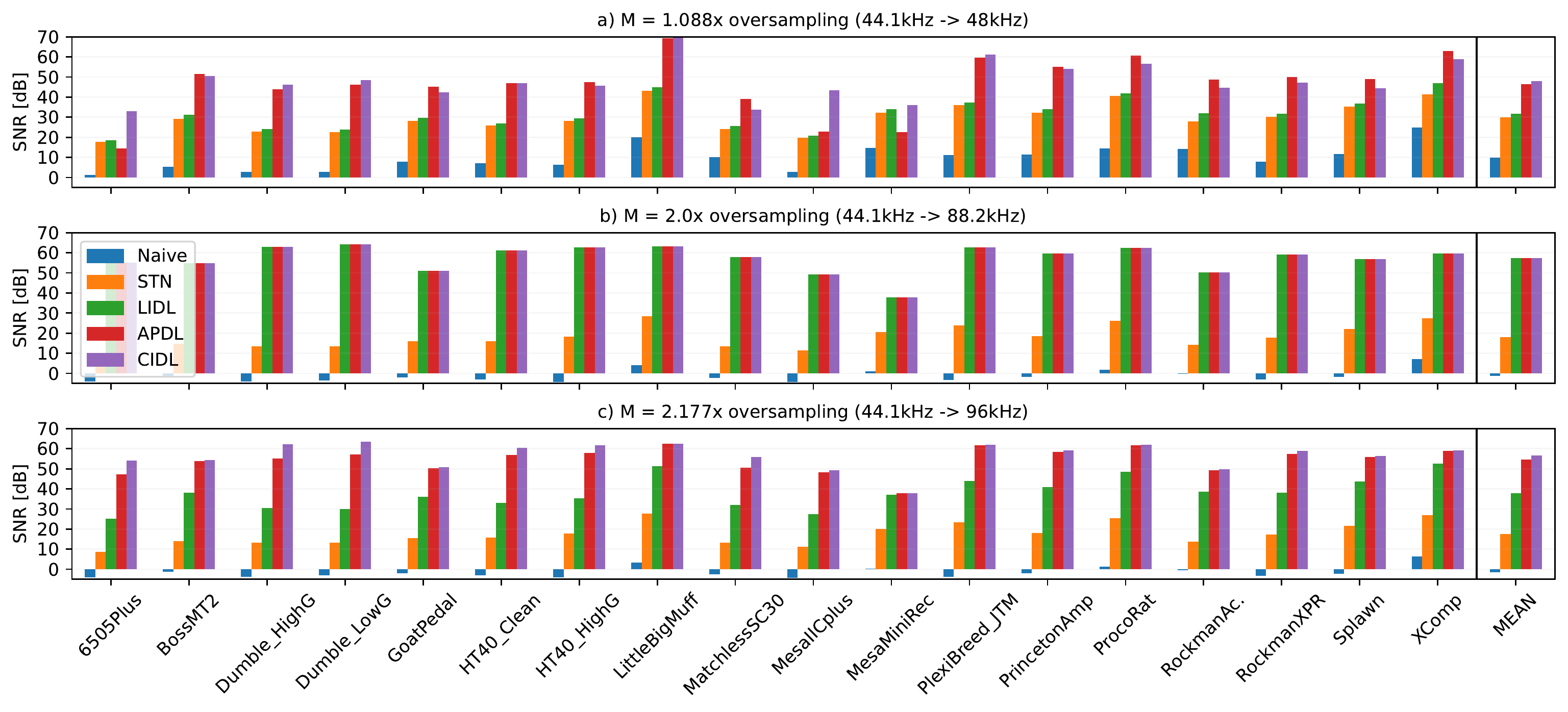}}
    \caption{\label{fig:audio_snr}{\it SNR results across LSTM models trained on various effects for common SR conversion ratios. The input audio signal was one minute of electric guitar and bass. The mean SNR over all effects is shown in the last column.}}
\end{figure*}
\begin{figure*}[h!]
    \centerline{\includegraphics[width=.99\textwidth, clip, trim=0mm 3mm 0mm 2mm]{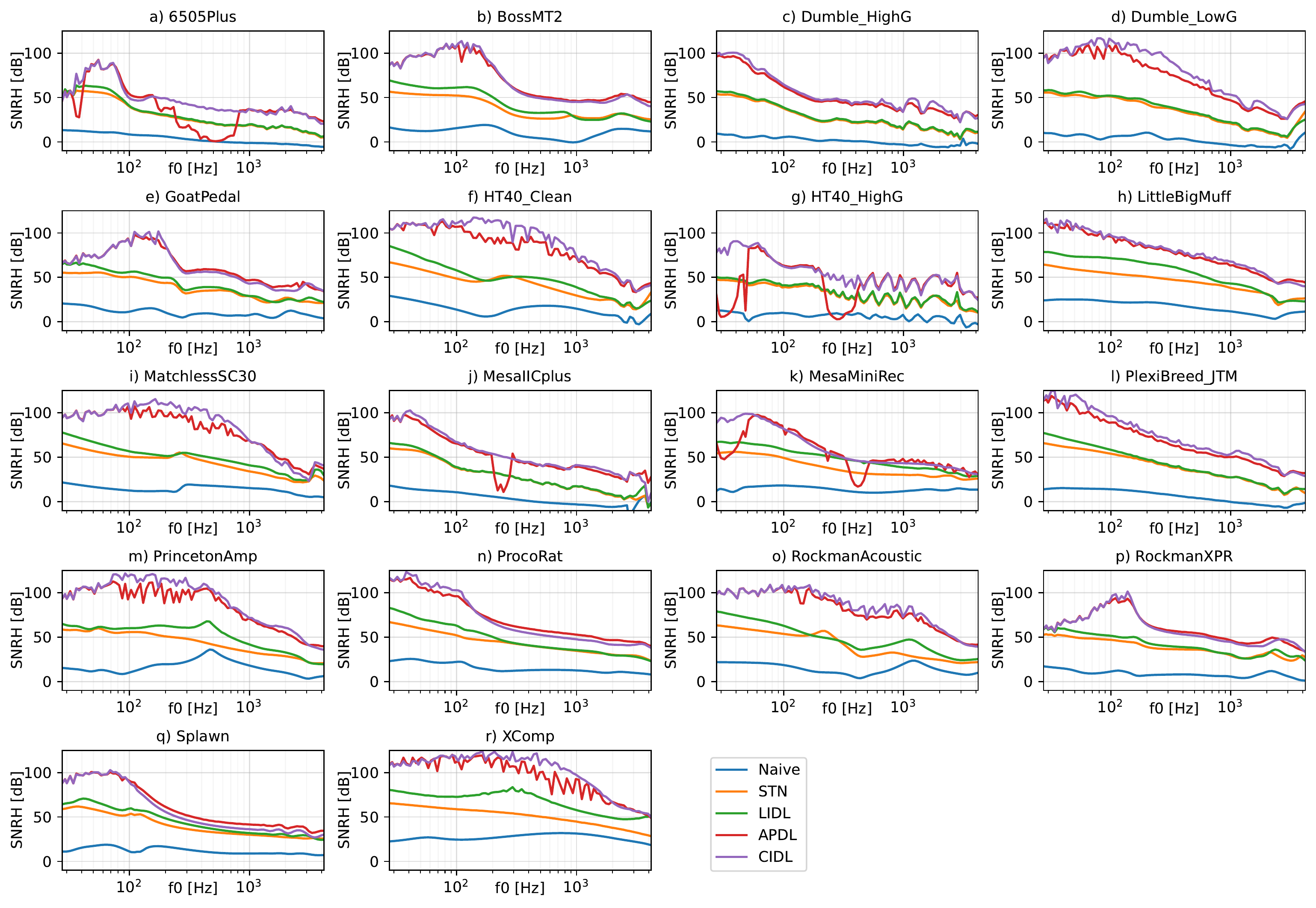}}
    \caption{\label{fig:sine_snrh}{\it Signal-to-noise ratio of harmonic components (SNRH) in the model output signal against input sine tone frequency, $f_0$, for LSTM models trained on various effects. The oversampling factor is $M=1.0884$ corresponding to SR modification from \SI{44.1}{\kHz} to \SI{48}{\kHz}.}}
\end{figure*}

\subsection{Test metrics and input signals}
To simulate a realistic use-case, the models were tested on one minute of guitar and bass recordings -- the same test set used in previous works \cite{Wright2020}. 
First we up-sampled the input signal to the target SR, $F'_s$, to obtain $\tilde{x}$ then passed this audio signal through the respective modified networks to obtain the output $\tilde{y}$. The output was then down-sampled to the base-rate ($\hat{y}$) and the signal-to-noise ratio (SNR) computed relative to the baseline model output at \SI{44.1}{\kHz}, $y$, using Eq. \eqref{eq:snr}. The up-sampling and down-sampling was implemented with the FFT-based method proposed in \cite{Valimaki2023} because of machine-precision level of accuracy.
\begin{eqnarray}\label{eq:snr}
    {\rm SNR} = \frac{\sum_{n=0}^{N-1} y^2}{\sum_{n=0}^{N-1} \left( y - \hat{y} \right)^2}.
\end{eqnarray}

For a more detailed analysis on how the proposed methods perform across a range of driving frequencies, and to evaluate aliasing reduction, we consider the case of sinusoidal inputs. Sine tones of duration one second were generated with frequencies ranging from $f_0 = $ 27.5\,Hz to 4186\,Hz (corresponding to the notes on a piano) at the base SR $F_s$ and the inference SR $F'_s$ respectively. The tones were then processed through the baseline model and the SR adjusted models to obtain $y$ and $\tilde{y}$ respectively. Two signal-to-noise metrics are then considered: signal to aliasing noise ratio (SNRA) and signal to harmonic noise ratio (SNRH). 

The SNRA is computed by first windowing $\tilde{y}$ with a $N'$-length Chebyshev window with 120\,dB stop-band attenuation and taking the DFT to get the spectrum $\tilde{Y}$. The amplitudes and phase of the harmonic components were extracted from the spectrum and used to synthesise an ideal alias-free version of the signal $\tilde{y}_{\rm BL}$ at the base SR. This signal was then windowed with an $N$-length Chebyshev window, transformed to obtain spectrum $\tilde{Y}_{\rm {BL}}$, and scaled such that the first harmonic of $\tilde{Y}_{\rm {BL}}$ and $\tilde{Y}$ were equal in amplitude. The SNRA was then computed as
\begin{eqnarray}
    {\rm SNRA} = \frac{\sum_{k=0}^{N/2}|\tilde{Y}_{\rm {BL}}[k]|^2}{\sum_{k=0}^{N/2} \left( |\tilde{Y}[k]| - |\tilde{Y}_{\rm {BL}}[k] |\right)^2},
\end{eqnarray}
where the $N/2$ bin corresponds to the Nyquist limit at the base SR $F_s$. To compare the non-aliased components of $y$ and $\tilde{y}$, a band-limited version of the baseline was synthesised, $y_{\rm BL}$, and the SNRH computed as
\begin{eqnarray}
    {\rm SNRH} = \frac
    {\sum_{n=0}^{N-1} y_{\rm BL}^2}{\sum_{n=0}^{N-1} \left( y_{\rm BL} - \tilde{y}_{\rm BL} \right)^2}.
\end{eqnarray}

\subsection{SNR Results}

Figure \ref{fig:audio_snr} shows the audio-input SNR results across the various LSTM models at common oversampling ratios (corresponding to SRs conversion from \SI{44.1}{\kHz} to \SI{48}{\kHz}, \SI{88.1}{\kHz} and \SI{96}{\kHz} respectively). Figure \ref{fig:sine_snrh} shows the results of the more detailed sine-tone-input analysis at \SI{48}{\kHz}. 

It is clear that in all cases, any of the methods outlined in Section \ref{sec:methods} provide a significant improvement in SNR compared to the the naive method of using the original (unmodified) RNN at arbitrary SRs. The STN method provides an increase in SNR, but gets progressively worse as the SR increases -- a trend also seen in the linear analysis. This suggests that the STN method should be avoided in practice for large SR changes. 

The delay-based methods -- LIDL, APDL and CIDL -- generally perform significantly better and do not deteriorate as SR increases. At $M=2$ times oversampling, all three methods reduce to a pure delay line and therefore give identical results. In this case the quality increase compared to the STN method is significant -- giving a minimum of 18\,dB (``MesaMiniRec'') and up to 50\,dB increase (``DumbleLowG'') in SNR. 

The results for non-integer oversampling ratios allow a comparison between the delay-based methods. When operating at \SI{48}{\kHz}, the LIDL method performs comparably to STN, and shows further improvement at \SI{96}{\kHz}. At \SI{48}{\kHz}, the APDL and CIDL methods generally give similar results apart from in three cases in which the APDL performs similar to or worse than STN and LIDL: ``6505Plus'', ``MesaIICplus'' and ``MesaMiniRec''. This is further demonstrated in Figure \ref{fig:sine_snrh}a), g), j) and k). 

In these models, and for certain sine tone inputs in the mid-frequency range, the APDL method fails catastrophically giving nearly 0\,dB SNRH.  
A more detailed examination in these failure cases showed oscillations at the Nyquist frequency and extreme aliasing that can perhaps be attributed to numerical round-off error due to pole-zero cancellation in the all-pass filter when the fractional delay is small ($\Delta < 0.1$) \cite{PASPWEB2010}. Further analysis is needed to determine why it is mid-frequency inputs that cause the problem, but this is a concerning result given that guitar and bass have harmonic content in this range. This problem motivated the investigation into higher order FIR interpolation, which is guaranteed stable. It can be seen that the CIDL method does not suffer from the anomalies seen in the APDL method and otherwise gives a comparable SNR.

For SR conversion from \SI{44.1}{\kHz} to \SI{96}{\kHz}, all delay-based methods improve in accuracy (compared to at \SI{48}{\kHz}), with CIDL providing the highest SNR across all models. The aforementioned anomalies in the APDL method results appear to be suppressed at this SR, and the APDL performs consistently second-best.

\subsection{Aliasing}
\begin{figure}
\centerline{\includegraphics[width=0.5\textwidth]{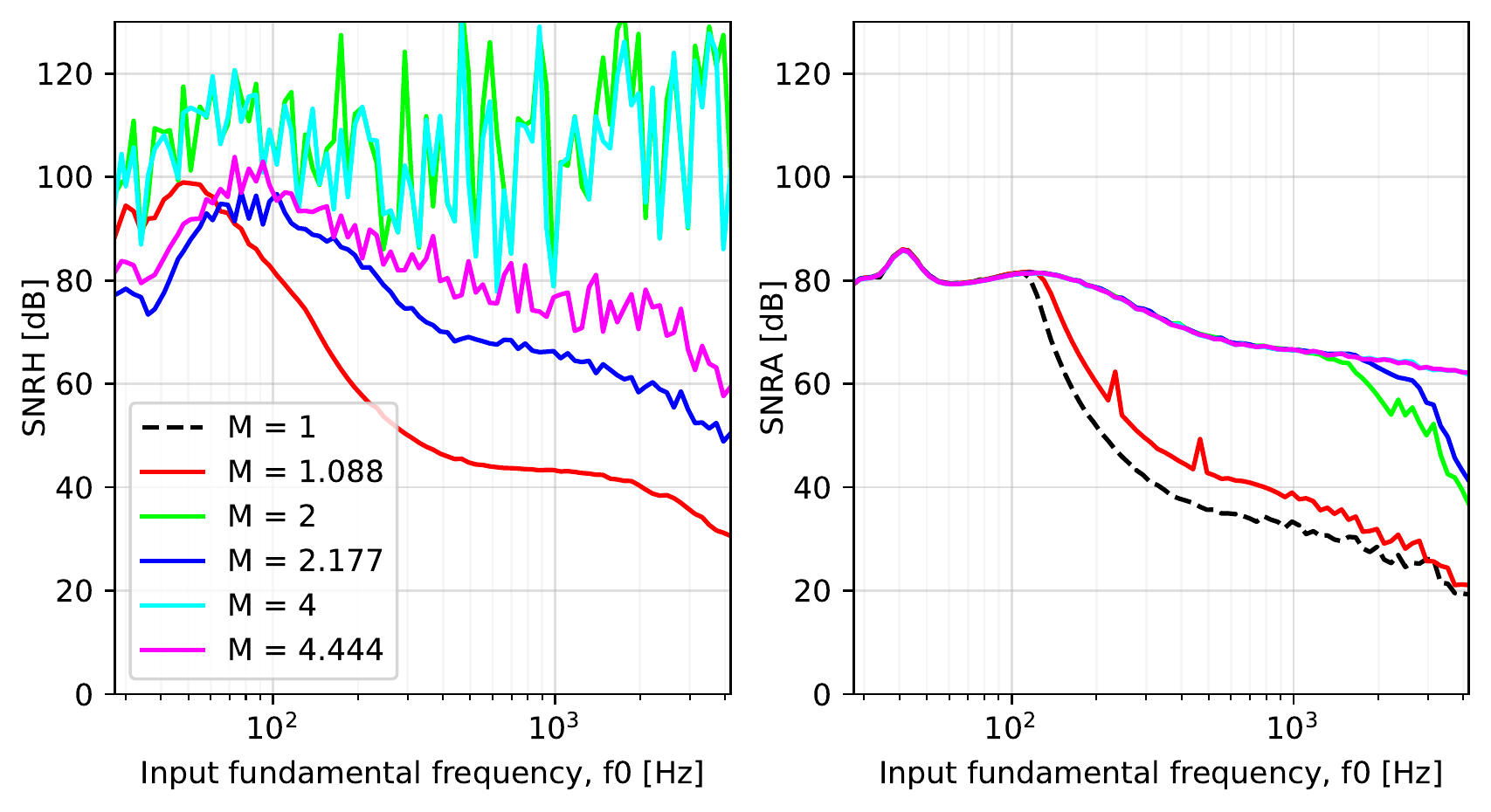}}
\caption{\label{fig:mesamini}{\it Signal-to-noise ratio of harmonics (left) and signal-to-aliasing-noise ratio (right) of the ``MesaMini'' model output with the proposed cubic interpolation method (CIDL) of SR conversion. Note that for $M=2, 4$ the CIDL method reduces to the pure delay-line method \cite{Chowdhury2022}. The baseline $M=1$ is not shown in LH figure because the SNRH is infinite.}}
\end{figure}

Figure \ref{fig:mesamini} shows how the proposed CIDL method performs across different oversampling ratios using the ``MesaMini'' model as a case study. This model was chosen as a worst-case-scenario because it exhibited the strongest aliasing at the base SR. In Figure \ref{fig:mesamini} (right), it can be seen that oversampling progressively reduces aliasing with the reduction most obvious for high-frequency input signals, as is to be expected. In Figure \ref{fig:mesamini} (left), we can see how oversampling affects the amplitudes of the expected harmonic components. It can be seen that for integer sampling ratios the SNRH is around 100\,dB -- meaning the harmonic amplitudes match the baseline to near machine precision (operating at single-precision). For non-integer ratios, the SNRH is lower due to interpolation error but generally increases as the SR increases. These results indicate that oversampling does indeed reduce aliasing in RNN models of guitar effects, and that if a high-quality interpolation method is used, such as Lagrange interpolation, the original harmonic distortion in the baseline model can be preserved to a high degree of accuracy.

\section{Conclusions and further work}
In this paper we examined methods of modifying recurrent neural network architectures to enable high-fidelity audio processing at a SR that differs from the training data. Integer and non-integer oversampling was considered with down-sampling left to future work. The methods were tested on several LSTM models of guitar effects and evaluated through the SNR of the modified (oversampled) model relative to the original model output.

Two distinct classes of method were compared: the STN method, in which the SR is adjusted by scaling the state residual (as in an explicit numerical solver); and delay-based methods, in which the length of the RNN sample delay is stretched by the oversampling factor. For non-integer oversampling, the latter method requires the estimation of a fractional delay. In this case we considered the previously proposed linear interpolation method (LIDL) and proposed, in addition, two alternative methods using a first-order all-pass filter and cubic Lagrange interpolation (APDL and CIDL, respectively).

For integer oversampling, the delay-based method gave consistently the highest fidelity across all models: improving the signal-to-aliasing noise ratio with little effect on the magnitude of harmonic components.

For non-integer oversampling, the proposed CIDL method provided a considerable improvement over the lower-order FIR LIDL method. This result suggests that higher-order Lagrange interpolation may improve results even further. The APDL method appeared promising, and for some models its performance was superior to that of the CIDL method, but crucially in some cases it fails and produces noisy, highly-aliased, inaccurate output. We therefore recommend the proposed CIDL method as the best option for non-integer SR conversion. 

\vspace{-5pt}
\label{sec:conc}
\section{Acknowledgments}

Part of this work was conducted when Alistair Carson visited Vesa Välimäki and Alec Wright at the Aalto Acoustics Lab on May 11 -- July 22, 2023. Stefan Bilbao also visited the Aalto Acoustics Lab on May 31--June 15, 2023.

\begin{footnotesize}
\bibliographystyle{IEEEbib}
\bibliography{DAFx24_tmpl}
\end{footnotesize}

\end{document}